\documentclass[twopage,11pt] {article}  

\begin{document}           

\title{Viscous Cosmology, Entropy, and the Cardy-Verlinde Formula}  
\author{Iver Brevik\footnote{Department of Energy and Process Engineering, Norwegian University of Science and Technology, N-7491 Trondheim, Norway.
E-mail address: iver.h.brevik@ntnu.no}   }         

\maketitle                 

\pagestyle{myheadings} \thispagestyle{plain} \markboth{}{A Sample
Document} \setcounter{page}{1}

\begin{abstract}

The holographic principle in a radiation dominated universe, as
discussed first by Verlinde [hep-th/008140], is extended so as to
incorporate the case of a bulk-viscous cosmic fluid. This
corresponds to a non-conformally invariant theory. Generalization
of the Cardy-Verlinde entropy formula to the viscous case appears
to be formally possible, although on physical grounds one may
question some elements in this type of theory, especially the
manner in which the Casimir energy is evaluated. Also, we consider
the observation made by Youm [Phys. Lett. {\bf B531}, 276 (2002)],
namely that the entropy of the universe is no longer expressible
in the conventional Cardy-Verlinde form if one relaxes the
radiation dominance equation of state for the fluid and instead
merely assumes that the pressure is proportional to the energy
density. We show that Youm's generalized entropy formula remains
valid when the cosmic fluid is no longer ideal, but endowed with a
constant bulk viscosity. In the introductory part of this article,
we take a rather general point of view and survey the essence of
cosmological theory applied to a fluid containing both a constant
shear viscosity and a constant bulk viscosity.

\end{abstract}

\section{Introduction}

Perfect fluid models have for a long time been used in
cosmological theory. The introduction of the viscosity concept
came later, and there has in fact been an open question as to
whether the viscosity concept is needed for an explanation of the
observed quantities in the universe. Misner \cite{misner68} was
probably the first to introduce viscosity in cosmology in
connection with his study of how initial anisotropies in the early
universe became relaxed. Cosmological model with viscosity have
later now and then been discussed in the literature, from various
points of view. A useful review of the subject, with many
references to the earlier literature up to 1990, was given by
Gr{\o}n \cite{gron90}.

From a physical point of view it would in our opinion be almost
surprising if the viscosity concept were {\it not} of importance
in cosmology. An essential ingredient of the Friedmannian model of
the universe is after all to borrow the energy-momentum tensor
from non-viscous fluid dynamics and insert it into the Einstein
equations. From fluid dynamics we know that in many situations the
non-viscous theory is inadequate, in particular, if anisotropies
or turbulence effects are involved. Use of the molecular shear and
bulk viscosities means in effect an expansion of the theory to
first order in the deviations from thermal equilibrium. Moreover,
once turbulence occurs in the velocity field it is the {\it
Reynolds stresses},  rather than the  stresses caused by molecular
viscosities, that govern the behavior of the fluid. In view of
these known facts from ordinary fluid dynamics it is natural to be
somewhat reluctant to regard cosmology as an exceptional case for
which the viscosity concepts are of no use. One main reason why
the viscous cosmological theory has so far not gained large
attention is undoubtedly that the viscosity-generated entropy
$\sigma$ per baryon is observed to be very large, $ \sigma \sim
10^9$, and previous investigations have shown that a
straightforward use of the kinematically derived bulk viscosity in
the isotropic and homogeneous universe is unable to explain this
large magnitude \cite{weinberg71,weinberg72,johri88}.

The fluidlike behavior of the universe during its early epochs is
in all probability very complicated. One particular facet of the
problem is the existence of one or more {\it phase transitions}
that may have led to sudden changes whose description is naturally
given in terms of the phenomenological viscosity coefficients. An
eclatant example of this sort is the transition from the de Sitter
universe back to the FRW universe at the end of the inflationary
era, at $t \sim t^{-33}$ s. We may call the effective viscosity
operative during such a phase transition ({\it bulk} viscosity in
case of an isotropic universe) an "impulsive" viscosity.  Examples
of this kind of theory were examined in
Refs.~\cite{brevik94,brevik96}.

Another interesting aspect of viscous cosmology that has arisen
recently is the influence from {\it shear} viscosity. The shear
viscosity comes into play in connection with anisotropy. Thus
Weinberg \cite{weinberg04} derived an analytic formula for the
traceless part of the anisotropic stress tensor due to freely
streaming neutrinos. Another recent example is the suggestion of
Kovtun et al. \cite{kovtun03} and Karch \cite{karch03} (cf. also
the discussion in \cite{brevik04}) that there exists in cosmology
a universal lower bound on the ratio $\eta /s$, where $\eta$ is
the shear viscosity and $s$ the entropy per unit volume.

The main topic to be dealt with in the present article is the
incorporation of the bulk viscosity $\zeta$ in the so-called
holographic principle for the early universe. In the case of a
radiation dominant ideal-fluid universe, Verlinde
\cite{verlinde00} put forward the idea that there exists a bound
on the subextensive entropy associated with the Casimir energy.
This bound is called the holographic bound. When this bound is
saturated, there exists a formal coincidence between the Friedmann
equation for $H^2 \equiv (\dot{a}/a)^2$ and the Cardy entropy
formula known from conformal field theory \cite{cardy86,blote86}.
The question that naturally arises is whether this merging between
the holographic principle, the entropy formula from conformal
field theory, and the Friedmann equation from cosmology, is of
deeper physical significance and thus reproducible under more
general conditions, or if it is just a formal coincidence. As one
would expect, the Verlinde proposal has been the subject of study
in cases where more general effects have been accounted for. For
instance, Wang {\it et al.} \cite{wang01} have considered
universes having a cosmological constant, and Nojiri {\it et al.}
\cite{nojiri01}  have considered quantum bounds for the
Cardy-Verlinde formula. There are several papers in this area of
Padmanabhan {\it et al.} \cite{padmanabhan87}, and of Cai {\it et
al.}  \cite{cai01}; the paper of Youm \cite{youm02} contains an
extensive list of references.  We also mention the considerable
interest that has arisen in connection with entropy and energy as
following from quantum and thermal fluctuations in conformal field
theories \cite{kutasov01}.

We begin in the next section by giving a survey of cosmological
theory applied to a fluid that is general enough to possess both a
constant shear viscosity $\eta$ and a constant bulk viscosity
$\zeta$. Thereafter, in Sect.~3 we consider the Verlinde setting
when the fluid is taken to possess a bulk viscosity only. In
Sect.~4 we carry out an analogous reasoning for the case that the
bulk-viscosity fluid is not necessarily radiation dominated but
instead required to obey the equation of state
\begin{equation}
p=(\gamma -1)\rho, \label{1}
\end{equation}
with $\gamma$ a constant. Sections 3 and 4 are based upon material
given previously  in Refs.~\cite{brevik02,brevik02a,brevik03}. We
supply the analysis with some numerical estimates; thus we exploit
the smallness of $\zeta$ in the plasma era to calculate the
viscosity-induced correction to the scale factor $a(t)$, assuming
for simplicity that the spatial curvature $k$ is zero. In Sect.~5
we test the proposed holographic entropy bound on an example,
taken from the plasma era in the early universe.

We mostly use natural units, with $\hbar =c=k_B=1$.

\section{Survey of viscous cosmology}

We use the convention in which the Minkowski metric is (-+++). Let
$U^\mu=(U^0,U^i)$ be the four-velocity of the cosmic fluid. In
comoving coordinates, $U^0=1,\,U^i=0$.

Let $g_{\mu\nu}$ be the general metric. Using the projection
tensor
\begin{equation}
h_{\mu\nu}=g_{\mu\nu}+U_\mu U_\nu , \label{2}
\end{equation}
we define the rotation tensor as
\begin{equation}
\omega_{\mu\nu}=h_\mu^\alpha h_\nu^\beta
U_{[\alpha;\beta]}=\frac{1}{2}(U_{\mu;\alpha}h_\nu^\alpha-U_{\nu;\alpha}h_\mu^\alpha)
\label{3}
\end{equation}
and the expansion tensor as
\begin{equation}
\theta_{\mu\nu}=h_\mu^\alpha h_\nu^\beta
U_{(\alpha;\beta)}=\frac{1}{2}(U_{\mu;\alpha}h_\nu^\alpha+U_{\nu;\alpha}h_\mu^\alpha).
\label{4}
\end{equation}
The scalar expansion is $\theta \equiv
\theta_\mu^\mu={U^\mu}_{;\mu}$. The shear tensor, as defined by
\begin{equation}
\sigma_{\mu\nu}=\theta_{\mu\nu}-\frac{1}{3}h_{\mu\nu}\,\theta,
\label{5}
\end{equation}
is traceless, $\sigma_\mu^\mu=0$. The following decomposition of
the covariant derivative is often useful:
\begin{equation}
U_{\mu;\nu}=\omega_{\mu\nu}+\sigma_{\mu\nu}+\frac{1}{3}h_{\mu\nu}\,
\theta-A_\mu U_\nu, \label{6}
\end{equation}
where $A_\mu \equiv \dot{U}_\mu=U^\nu U_{\mu;\nu}$ is the
four-acceleration of the fluid.

We write down the expression for the energy-momentum tensor
$T_{\mu\nu}$ of the viscous fluid, taking into account also the
conduction of heat. As already mentioned, $\eta$ and $\zeta$ are
respectively the shear and the bulk viscosities;  if moreover
$\kappa$ is the thermal conductivity (all quantities taken in
accordance with their nonrelativistic definitions), then
\begin{equation}
Q^\mu=-\kappa h^{\mu\nu}(T_{,\nu}+TA_\nu) \label{7}
\end{equation}
is the spacelike heat flux density four-vector, and
\begin{equation}
T_{\mu\nu}=\rho U_\mu U_\nu+(p-\zeta \theta)h_{\mu\nu}-2\eta
\sigma_{\mu\nu}+Q_\mu U_\nu+Q_\nu U_\mu. \label{8}
\end{equation}
Here $\rho$ is the mass-energy density and $p$ is the isotropic
pressure, both taken in the local rest inertial frame. The last
term in Eq.~(\ref{7}), containing $TA_\nu$, is of relativistic
origin. If one ignores this term, one is left with $Q^\mu=-\kappa
h^{\mu\nu}T_{,\nu}$. This expression is defined such that in a
local rest inertial frame (designated with a "hat")
$Q_{\hat{0}}=0$, whereas $Q_{\hat{i}}=-\kappa T_{,\hat{i}}$ is the
heat energy per unit time crossing a unit surface orthogonal to
the unit vector ${\bf{e}}_{\hat{i}}$.

Consider next the production of entropy. It is here instructive to
start from nonrelativistic theory. If $u_i$ are the
nonrelativistic velocity components, and $\sigma$ the
nonrelativistic entropy per particle (baryon), then the ordinary
entropy per unit volume is $S=n\sigma$, $n$ being the baryoun
number density. From nonrelativistic theory we have
\cite{landau87}
\begin{equation}
\frac{dS}{dt}=\frac{2\eta}{T}(\theta_{ik}-\frac{1}{3}\delta_{ik}{\bf
\nabla \cdot u})^2+\frac{\zeta}{T}({\bf \nabla \cdot u})^2
+\frac{\kappa}{T^2}(\nabla T)^2, \label{9}
\end{equation}
where $\theta =u_{(i,k)}$. The transition to relativistic theory
may be made via the effective substitutions
\begin{equation}
\theta_{ik} \rightarrow \theta_{\mu\nu}, \quad \delta_{ik}
\rightarrow h_{\mu\nu}, \quad {\bf \nabla \cdot u}\rightarrow
\theta, \quad -\kappa T_{,k}\rightarrow Q_\mu, \label{10}
\end{equation}
from which we obtain
\begin{equation}
{S^\mu}_{;\mu}=\frac{2\eta}{T}\sigma_{\mu\nu}\sigma^{\mu\nu}+\frac{\zeta}{T}\theta^2+\frac{1}{\kappa
T^2}Q_\mu Q^\mu. \label{11}
\end{equation}
Here, $S^\mu$ is the entropy four-vector
\begin{equation}
S^\mu=n\sigma U^\mu+\frac{1}{T}Q^\mu. \label{12}
\end{equation}
Our treatment above follows Ref.~\cite{brevik94}. The same result
is obtained from a more careful analysis taking into account the
relativistic thermodynamic equations \cite{weinberg71,taub78}.

In the case of thermal equilibrium,  $Q_\mu=0$. Moreover, in
accordance with usual practice we omit the shear viscosity in view
of the assumed complete isotropy of the cosmic fluid, although we
have to remark that this is actually a nontrivial point. The
reason is that the shear viscosity is usually so much greater than
the bulk viscosity. Typically, after termination of the plasma era
at the time of recombination ($T \simeq 4000$ K), the ratio $\eta
/\zeta$ as calculated from kinetic theory is as large as about
$10^{12}$ \cite{brevik94}. Thus, even a slight anisotropy in the
fluid would easily outweigh the effect of the effect of the very
small bulk viscosity.

\section{Cardy-Verlinde formula in a bulk-viscous radiation
dominated universe}

Assume now that the metric is of the Friedmann-Robertson-Walker
(FRW) type,
\begin{equation}
ds^2=-dt^2+a^2(t)\left( \frac{dr^2}{1-kr^2}+r^2 d\Omega^2
\right),\label{13}
\end{equation}
$k=-1,0,1$ being the curvature parameter. In comoving coordinates,
$U_{\mu ;\nu}=\Gamma_{\mu\nu} ^0$, so that in view of the standard
relations for the Christoffel symbols we get
${U^\mu}_{;\nu}=h_\nu^\mu \,\dot{a}/a$, with $\dot{a}=da/dt$. From
these equations we see that the rotation and shear tensors both
vanish,
\begin{equation}
\omega_{\mu\nu}=\sigma_{\mu\nu}=0, \label{14}
\end{equation}
whereas the scalar expansion is
\begin{equation}
\theta=3\dot{a}/a=3H, \label{15}
\end{equation}
$H$ being the Hubble parameter. The four-acceleration is zero,
$A_\mu=0$. The energy-momentum tensor becomes
\begin{equation}
T_{00}=\rho, \quad T_{0k}=0, \quad T_{ik}=(p-\zeta \theta)g_{ik}.
\label{16}
\end{equation}
There is thus no conduction of heat in the FRW space, which is a
homogeneous space. The entire effect of the bulk viscosity is to
reduce the pressure by an amount $\zeta \theta$, so that the
effective pressure becomes $\tilde{p} = p-\zeta \theta $.

When applied to the FRW space, Eqs.~ (\ref{11}) and  (\ref{12})
yield
\begin{equation}
{S^\mu}_{;\mu}=\frac{\zeta}{T}\theta^2,\quad S^0=n\sigma, \quad
S^i=0. \label{17}
\end{equation}
Consider now the Einstein equations
 \begin{equation}
R_{\mu\nu}-\frac{1}{2}Rg_{\mu\nu}+\Lambda g_{\mu\nu}=8\pi
G\,T_{\mu\nu}, \label{18}
 \end{equation}
 from which we  obtain the first Friedmann
equation (the "initial value equation")
\begin{equation}
H^2=\frac{8\pi
G}{3}\,\rho+\frac{\Lambda}{3}-\frac{k}{a^2},\label{19}
\end{equation}
$\rho=E/V$ being the  energy density. This equation contains no
viscous term. The second Friedmann equation (the "dynamic
equation"), when combined with Eq.~(\ref{19}), yields
\begin{equation}
\dot{H}=-4\pi G (\rho +\tilde{p})+\frac{k}{a^2},\label{20}
\end{equation}
in which the presence of viscosity is explicit.

The differential conservation for energy, ${T^{0\nu}}_{;\nu}=0$,
yields
\begin{equation}
\dot{\rho}+(\rho +p)\theta =\zeta \theta^2. \label{20a}
\end{equation}
The conservation equation for baryon particle number is
\begin{equation}
(nU^\mu)_{;\mu}=0, \label{20b}
\end{equation}
which means that $na^3=$ constant in the comoving frame. Then from
Eq.~(\ref{17}) we obtain, when substituting $S^\mu$ from
Eq.~(\ref{12}) and observing that $Q^0=0$ in the comoving frame,
\begin{equation}
n\dot{\sigma}=\frac{\zeta}{T}\theta^2. \label{20c}
\end{equation}

We now recall that the entropy of a (1+1) dimensional CFT is given
by the Cardy formula \cite{cardy86, blote86}
\begin{equation}
S=2\pi \sqrt{\frac{c}{6}\left(L_0-\frac{c}{24}\right)},\label{21}
\end{equation}
where $c$ is the central charge and $L_0$ the lowest Virasoro
generator.

  Let us assume that the universe is closed, and has a vanishing cosmological constant, $k=+1,~~~\Lambda=0$.
This is the case considered by Verlinde  \cite{verlinde00} (in his
formalism the number $n$ of space dimensions is set equal to 3).
The Friedmann equation (\ref{19}) is seen to agree with the CFT
equation (\ref{21}) if we perform the substitutions
\begin{equation}
 L_0 \rightarrow \frac{1}{3}Ea,~~~c\rightarrow \frac{3}{\pi}\frac{V}{Ga},~~~S\rightarrow
 \frac{HV}{2G}.\label{22}
\end{equation}
 These substitutions are the same as in Ref.\cite{verlinde00}. We
see thus that Verlinde's argument remains valid, even if the fluid
possesses a bulk viscosity. Note that no assumptions have so far
been made about the equation of state for the fluid. At this point
the following question thus naturally arises: is there a deeper
connection between the laws of general relativity and those of
quantum field theory?

Continuing this kind of reasoning, let us consider the three
actual entropy definitions.

\begin{itemize}

\item First, there is
the Bekenstein entropy \cite{bekenstein81},
\begin{equation}
S_B=\frac{2\pi}{3}Ea. \label{23}
\end{equation}
The arguments for deriving this expression seem to be of a general
nature; in accordance with Verlinde we find it likely that the
Bekenstein bound $S \leq S_B$ is universal. We shall accept this
expression for $S_B$ in the following, even when the fluid is
viscous.

\item

The next kind of entropy is the Bekenstein-Hawking expression
$S_{BH}$, which is supposed to hold for systems with limited
self-gravity:
\begin{equation}
 S_{BH}=\frac{V}{2Ga}. \label{24}
 \end{equation}
Again, this expression relies upon the viscous-insensitive member
(\ref{19}) of Friedmann's equations. Namely, when $\Lambda=0$ this
equation yields
\[ S_B < S_{BH} \quad {\rm when} \quad Ha < 1, \]
\begin{equation}
 S_B > S_{BH} \quad {\rm when} \quad Ha > 1. \label{25}
\end{equation}
The borderline case between a weakly and a strongly gravitating
system is thus at $Ha=1$. It is reasonable to identify $S_{BH}$
with the holographic entropy of a black hole with the size of the
universe.

\item

The third entropy concept is the Hubble entropy $S_H$. It can be
introduced by starting from the conventional formula $A/4G$ for
the entropy of a black hole. The horizon area $A$ is approximately
$H^{-2}$, so that
\begin{equation}
S_H \sim \frac{H^{-2}}{4G} \sim \frac{HV}{4G}, \label{26}
\end{equation}
 since $V\sim H^{-3}$. Arguments
have been given by Easther and Lowe \cite{easther99}, Veneziano
\cite{veneziano99}, Bak and Rey \cite{bak00}, and Kaloper and
Linde \cite{kaloper99} for assuming the maximum entropy inside the
universe to be produced by black holes of the size of the Hubble
radius (cf. also \cite{fischler98}). According to Verlinde the FSB
prescription (see \cite{verlinde00} for a closer discussion) one
can determine the prefactor:
\begin{equation}
S_H=\frac{HV}{2G}. \label{27}
\end{equation}
It is seen to agree with Eq.~(\ref{22}).

\end{itemize}

One may now {\it choose} (see below) to define the Casimir energy
$E_C$ as the violation of the Euler identity:
\begin{equation}
E_C = 3(E+pV-TS) \label{28}
\end{equation}
 where, from scaling, the total energy $E$ can be decomposed as
($E_E$ is the extensive part)
$E(S,V)=E_E(S,V)+\frac{1}{2}E_C(S,V)$. Due to  conformal
invariance the products $E_E\, a$ and $E_C\, a$ are independent of
the volume $V$, and a function of the entropy $S$ only. From the
known extensive behaviour of $E_E$ and the sub-extensive behaviour
of $E_C$ one may write (for CFT)
 \begin{equation}
E_E=\frac{\alpha}{4\pi a} S^{4/3},~~~~E_C=\frac{\beta}{2\pi a}
S^{2/3}, \label{29}
\end{equation}
where $\alpha, \beta $ are constants whose product for CFTs is
known: $\sqrt{\alpha \beta}=n=3$ (this follows from the AdS/CFT
correspondence, cf. \cite{verlinde00}). From these expressions it
follows that
\begin{equation}
S=\frac{2\pi a}{3}\sqrt{E_C(2E-E_C)}. \label{30}
\end{equation}
This is the Cardy-Verlinde formula for the radiation dominated
universe. Identifying $Ea$ with $L_0$ and $E_C\,a$ with $c/12$ we
see that Eq.~(\ref{30}) becomes the same as Eq.~(\ref{21}), except
from a numerical prefactor which is related to our assumption
about $n=3$ space dimensions instead of $n=1$ as assumed in the
Cardy formula.

The question is now: can the above line of arguments be carried
over to the case of a viscous fluid? The most delicate point here
appears to be the assumed pure entropy dependence of the product
$Ea$. As we mentioned above, this property was derived from
conformal invariance, a property that is absent in the case under
discussion. To examine whether the property still holds when the
fluid is viscous (and conformal invariance is lost), we can start
from the Friedmann equations (\ref{19}) and (\ref{20}), in the
case $k=1, \, \Lambda=0$, and derive the "energy equation", which
can be transformed to
\begin{equation}
\frac{d}{da}(\rho a^4)=(\rho -3\tilde{p})a^3. \label{31}
\end{equation}
Thus, for a radiation dominated universe, $p=\rho /3$, it follows
that
\begin{equation}
\frac{d}{dt}(\rho a^4)=\zeta \,\theta^2 a^4. \label{32}
\end{equation}
Let us compare this expression, which is essentially the time
derivative of the volume density of the quantity $Ea$ under
discussion, with our earlier expression (\ref{20c}): both the two
time derivatives are seen to be proportional to $\zeta$. Since
$\zeta$ is small, we can insert for $a=a(t)$ the expression
pertinent for a non-viscous, closed universe: $a(t)=a_* \sin
\eta$, where $\eta$ here denotes conformal time, and $a_*$ is the
constant
\begin{equation}
 a_* =\sqrt{(8\pi G/3)\rho_{in} a_{in}^4}. \label{33}
 \end{equation}
The subscript "in"  designates the initial instant of the onset of
viscosity.  Imagine now that Eqs.~(\ref{32}) and (\ref{20c}) are
integrated with respect to time. Then, since the densities
$\zeta^{-1}\rho a^4$ and $\zeta^{-1} n\sigma$ can be drawn as
functions of $t$, it follows that $\rho a^4$ can be considered as
a function of $n\sigma$, or, equivalently, that $Ea$ can be
considered as a function of $S$. We conclude that this property,
previously derived on the basis of CFT, really appears to carry
over to the viscous case.

The following point ought to be commented upon. The specific
entropy $\sigma$ in Eq.~(\ref{20c}) is the usual thermodynamic
entropy per particle. The identification of $S$ with $HV/(2G)$, as
made in Eq.~(\ref{22}), is however something different, since it
is derived from a comparison with the Cardy formula (\ref{21}).
Since this entropy is the same as the Hubble entropy $S_H$ we can
write the equation as $ n\sigma_H=H/(2G), $ where  $\sigma_H$ is
the Hubble entropy per particle. This quantity is different from
$\sigma$, since it does not follow from thermodynamics plus
Friedmann equations alone, but from the holographic principle. The
situation is actually not peculiar to viscous cosmology. It occurs
if $\zeta=0$ also. The latter case is easy to analyze
analytically, if we focus attention on the case $t\rightarrow 0$.
Then, for any value of $k$, we have $a \propto t^{1/2}$, implying
that $H=1/(2t)$. Moreover, the equation of continuity
$(nU^\mu)_{;\mu}=0$ implies, as we have seen,   that
 $na^3=$ constant for a FRW universe, so that $n \propto t^{-3/2}$. The above
equation for $\sigma_H$  then yields
    $\sigma_H \propto t^{1/2}$. This is obviously different from
    the result for the thermodynamic entropy $\sigma$: from Eq.~(\ref{20c})
    we simply get $\sigma=$ constant when $\zeta=0$.
    The two specific entropies are thus different even in this case.

\subsection{Remarks on the physical interpretation}

   Let us make three remarks on the interpretation of the
above formalism. They are based on physical, rather than
mathematical, considerations.

\begin{itemize}

\item

 First, one may wonder about the
legitimacy of defining the Casimir energy such as in
Eq.~(\ref{28}). Usually, within the Green function approach, in a
spherical geometry the Casimir energy is calculated indirectly, by
integration of the Casimir surface force density $f=-(1/4\pi
a^2)\partial E/\partial a$. The force $f$ in turn is calculated by
first subtracting off the volume-dependent parts of the two scalar
Green functions; this is in agreement with the physical
requirement that $f \rightarrow 0$ at $r \rightarrow \infty$. (The
typical example of this configuration is that of a conducting
shell; cf., for instance, Ref.~\cite{milton78}.)
 That this kind of procedure should lead to the same result as Eq.~(\ref{28}),
which merely expresses a violation of the thermodynamic Euler
identity, is in our opinion not evident.

\item

Our second remark is about the physical meaning of taking the
Casimir energy $E_C$ to be {\it positive}. Verlinde assumes that
$E_C$ is bounded by the total energy $E$: $ E_C \leq E.$ This may
be a realistic bound for some of the CFTs. However, in general
cases, it is not true. For a realistic dielectric material it is
known that the full Casimir energy is not positive; the dominant
terms in $E_C$ are definitely {\it negative}. From a statistical
mechanical point of view this follows immediately from the fact
that the Casimir force is the integrated effect of the attractive
van der Waals force between the molecules. Now the case of a
singular conducting shell is  complicated - there are two limits
involved, namely the infinitesimal thickness of the shell and also
the infinite conductivity (or infinite permittivity) - and a
microscopical treatment of such a configuration has to our
knowledge not been given. What {\it is} known, is the
microscopical theory for a dielectric ball. Let us write down, for
illustration, the expression derived by Barton \cite{barton99} for
a dilute ball:
\[ E_C=-\frac{3\gamma}{2\pi^2}\frac{V}{\lambda^4} \]
\begin{equation}
+\gamma^2 \left( -\frac{3}{128
\pi^2}\frac{V}{\lambda^4}+\frac{7}{360\pi^3}\frac{A}{\lambda^3}
-\frac{1}{20\pi^2}\frac{1}{\lambda}+\frac{23}{1536\pi}\frac{1}{a}
\right), \label{34}
\end{equation}
where $\gamma =(\epsilon-1)/\epsilon,~~A$ is the surface area, and
$\lambda$ is a cutoff parameter. This expression, derived from
quantum mechanical perturbation theory, agrees with the
statistical mechanical calculation in Ref.~\cite{hoye00}, and also
essentially with Ref.~\cite{bordag99} (there are some numerical
factors different in the cutoff-dependent terms). It is evident
from this expression that the dominant, cutoff, dependent voume
terms, are negative.

We see that there remains one single, cutoff independent, term in
Eq.~(\ref{34}). This term is in fact positive. It can be derived
from macroscopic electrodynamics also,  by using either
dimensional continuation or zeta-function regularization, as has
been done in Ref.~\cite{brevik99}. In the present context the
following question becomes however natural: how can a positive,
small, cutoff dependent term in the Casimir energy play a major
role in cosmology? In another words, why should the matter
necessarily be conformal? Of course, our universe is different
from a dielectric ball, and we are not simply stating that
Verlinde's method is incorrect.
 Our aim is merely to stress the need of caution, when  results from one field in physics are applied to another field.
In any case, all this suggests that the consideration of
non-conformally invariant situations should significally change
the dynamical entropy bounds and bounds for Casimir energy.

\item

Our discussion on the generalized Verlinde formula in  the present
section was based upon the set of cosmological assumptions $ \{
p=\rho/3,\, k=+1, \,\Lambda=0 \}.$ The recent development of Wang
et al. \cite{wang01} is interesting, since it allows for a
nonvanishing cosmological constant (still assuming a closed
model). One of the scenarios treated in \cite{wang01} is that of a
de Sitter universe ($\Lambda
>0$) occupied by a universe-sized black hole. A black hole in de
Sitter space has the metric
\begin{equation}
 ds^2=-f(r)dt^2+f^{-1}(r)dr^2+r^2 d\Omega^2, \label{35}
 \end{equation}
where $f(r)=1-2MG/r-\Lambda r^2/3$. The region of physical
interest is that lying between the inner black hole horizon and
the outer cosmological horizon, the latter being determined by the
magnitude of $\Lambda$.

Although we do not enter into any detail about this theory, we
make the following observations: the above metric is {\it static};
there is no time-dependent scale factor involved, and the
influence from viscosity will not turn up in the line element.
Moreover, Wang et al. make use of only the member (\ref{19}) of
Friedmann's equations which, as we have noticed, is formally
independent of viscosity.

Does this imply that viscosity is without any importance for the
present kind of theory? The answer in our opininon is no, since
the theory operates implicitly with the concept of the maximum
scale factor $a_{max}$ in the closed Friedmann universe. In order
to calculate $a_{max}$, one has to solve the Friedmann equation
(\ref{20}) also, which contains the viscosity through the modified
pressure $\tilde{p}$. Thus, viscosity comes into play after all,
though in an indirect way.

\end{itemize}

\section{Cardy-Verlinde formula in the presence of a general
equation of state}

The paper of Youm mentioned above \cite{youm02} is interesting,
since it shows that the entropy of the universe can no longer be
expressed in the conventional form of Eq.~(\ref{30}) if one
relaxes the radiation dominance state equation and instead assumes
the more general equation (\ref{1}). And this brings us naturally
to the problem of how the presence of a bulk viscosity, together
with Eq.~(\ref{1}), influences the entropy formula. Actually, it
will turn out that the modified entropy formula found by Youm
still persists, even when a constant bulk viscosity is allowed
for. The central formula is Eq.~(\ref{42} ) below.

The main part of the formalism has been given already. Thus
Friedmann's equations take the same form (\ref{19}) and (\ref{20})
as before (with $k=+1, \Lambda =0$), the conservation equation for
energy is still Eq.~(\ref{20a}), and also the production rate for
entropy is as in Eq.~(\ref{20c}). However, Eq.~(\ref{32}) gets
replaced by
\begin{equation}
\frac{d}{dt}(\rho a^{3\gamma})=\zeta \theta^2 a^{3\gamma}.
\label{36}
\end{equation}

We can now carry out the same kind of reasoning as above: Since
$\zeta$ is assumed small, we can use for $a=a(t)$ the same
expression as for a nonviscous closed universe, namely $a(t)=a_*
\sin \eta$ with $a_*$ given by Eq.~(\ref{33})  (although this
approximate expression strictly speaking assumes that the universe
is radiation dominated). Imagine that Eqs.~(\ref{36}) and
(\ref{20c}) are integrated with respect to time. Since
$\zeta^{-1}\rho a^{3\gamma}$ and $\zeta^{-1}n\sigma $ can be drawn
as functions of $t$, it follows that $\rho a^{3\gamma}$ can be
considered as a function of $n\sigma$. Then, since the total
energy is $E \sim \rho a^3$ and the total entropy is $S \sim
n\sigma a^3$, it follows that $E a^{3(\gamma-1)}$ is independent
of the volume $V$ and is a function of $S$ only. This generalizes
the pure entropy dependence of the product $Ea$, found by Verlinde
\cite{verlinde00} in the case of a nonviscous radiation dominated
universe. And it is noteworthy that the derived property of
$Ea^{3(\gamma-1)}$ formally agrees exactly with the property found
by Youm \cite{youm02} when $\zeta=0$.

Let us carry out the analysis a bit further, and write the total
energy $E$ as a sum of an extensive part $E_E$ and a subextensive
part $E_C$, as in the previous section.
 Under a scale transformation $S \rightarrow \lambda S$
and $ V \rightarrow \lambda V $ with constant $\lambda$, $E_E$
scales linearly with $\lambda$. But the term $E_C$ scales with a
power of $\lambda$ that is less than one: as $E_C$ is the volume
integral over a local energy density expressed in the metric and
its derivatives, each of which scales as $\lambda^{-1/3}$, and as
the derivatives occur in pairs, the power in $\lambda$ has to be
1-2/3= 1/3. Thus we have
\begin{equation}
E_E(\lambda S, \lambda V)=\lambda E_E(S,V), \quad E_C(\lambda
S,\lambda V)=\lambda^{1/3}\,E_C(S,V), \label{40}
\end{equation}
which implies
\begin{equation}
E_E=\frac{\alpha}{4\pi a^{3(\gamma-1)}}\,S^\gamma,\quad
E_C=\frac{\beta}{2\pi a^{3(\gamma-1)}}\,S^{\gamma-2/3}. \label{41}
\end{equation}
Together with the decomposition of $E(S,V)$, this leads to
\begin{equation}
S=\left[ \frac{2\pi a^{3(\gamma-1)}}{\sqrt{\alpha
\beta}}\sqrt{E_C(2E-E_C)}\right]^{\frac{3}{3\gamma-1}}. \label{42}
\end{equation}
 This is the generalized Cardy-Verlinde formula, in agreement
with Eq.~(20) in Youm's paper, reducing to the standard formula
(with square root) in the case of a radiation dominated universe.
In conclusion, we have extended the basis of Eq.~(\ref{42}) so as
to include the presence of a constant bulk viscosity in the cosmic
fluid.

\subsection{Numerical estimates for a spatially flat universe}

It is of physical interest to supplement the above considerations
with some simple numerical estimates, showing, in particular, the
order of magnitude of the viscosity terms. Let us assume, as
mentioned above,  that the effect of the bulk viscosity
effectively sets in at some initial instant $t=t_{\rm in}$ and
that $\zeta$ is thereafter constant for the times that we
consider. For definiteness we choose
\begin{equation}
t_{\rm in}=1000~{\rm s} \label{43}
\end{equation}
after the big bang. The universe is then in the plasma (or
radiation) era; it consists of ionized H and He in equilibrium
with radiation. The particle density is $n_{\rm in} \simeq
10^{19}~{\rm cm^{-3}}$, and the temperature is $T_{\rm in} \simeq
4\times 10^8$ K. The advantage of considering this relatively late
stage of the universe's history is that the magnitude of $\zeta$
can be calculated using conventional kinetic theory. At $t=t_{\rm
in}$ one finds
\begin{equation}
\zeta=7.0\times 10^{-3}~{\rm g\,cm^{-1}\, s^{-1}} \label{44}
\end{equation}
(cf. \cite{brevik94} and further references therein).

For our estimate purposes it appears reasonable to assume that the
influence from the spatial curvature is not very important. For
simplicity let us put $k=0$. This implies that there exists the
following simple differential equation for the scalar expansion
\cite{brevik94}:
\begin{equation}
\dot{\theta}(t)+\frac{1}{2}\gamma \,\theta^2(t)-12\pi G \zeta
\theta(t)=0, \label{45}
\end{equation}
which can be solved to give the expression for the scale factor:
\begin{equation}
a(t)=a_{\rm in}\left[ 1+\frac{1}{2}\gamma\, \theta_{\rm in}
t_c\left( e^{(t-t_{\rm in})/t_c}-1\right)
\right]^{2/(3\gamma)},\label{46}
\end{equation}
where
\begin{equation}
t_c=(12\pi G\zeta)^{-1}. \label{47}
\end{equation}
The corresponding expression when viscosity is absent, is
\begin{equation}
a(t,\zeta=0)=a_{\rm in}\left[ 1+\frac{1}{2}\gamma \theta_{\rm
in}(t-t_{\rm in})\right]^{2/(3\gamma)}. \label{48}
\end{equation}
The ratio between the expressions (\ref{46}) and (\ref{48})
reduces to unity in the limit when $(t-t_{\rm in})/t_c \ll 1$.
This is we would expect. The influence from viscosity generally
turns up only in the factor $t_c$, and the effect becomes
strengthened when $t_c$ becomes smaller, {\it i.e.}, when $\zeta$
becomes larger. The numerical values given above for the instant
$t_{\rm in}=1000$ s correspond to
\begin{equation}
\theta_{\rm in}=1.5\times 10^{-3}~{\rm s^{-1}},\quad t_c=5.1\times
10^{28}~{\rm s}. \label{49}
\end{equation}

\subsection{ Perturbative expansion for a radiation dominated closed universe}

The smallness of $\zeta$ makes it natural, as an alternative to
the approach of the previous subsection, to make a Stokes
expansion in $\zeta$. For simplicity we now assume that $\gamma
=4/3$, {\it i.e.}, that the universe is radiation dominated. We
put $k=1.$ Let subscript zero refer to quantities in the
nonviscous case. We write the first order expansions as
\begin{equation}
a=a_0(1+\zeta a_1), \quad \rho=\rho_0 (1+\zeta \rho_1), \label{50}
\end{equation}
where the functions $a_1$ and $\rho_1$ are of zeroth order in
$\zeta$; they are regarded as functions of $t$ or alternatively as
functions of the conformal time $\eta$. Correspondingly, the
scalar expansion $\theta =3\dot{a}/a$ is written as
\begin{equation}
\theta=\theta_0(1+\zeta \theta_1). \label{51}
\end{equation}
The Friedmann equation (\ref{19}), and Eq.~(\ref{36}), now yield
to the zeroth order
\begin{equation}
\theta_0^2=24\pi G\rho_0-\frac{9}{a_0^2}, \label{52}
\end{equation}
\begin{equation}
\rho_{\rm in}a_{\rm in}^4=\rho_0 a_0^4, \label{53}
\end{equation}
and to the first order
\begin{equation}
\theta_0^2\,\theta_1=12\pi
G\rho_0\rho_1+\frac{9a_1}{a_0^2},\label{54}
\end{equation}
\begin{equation}
\rho_{\rm in}a_{\rm in}^4(\dot{\rho}_1+4\dot{a}_1)=\theta_0^2\,
 a_0^4. \label{55}
\end{equation}
The solutions of Eqs.~(\ref{52}) and (\ref{53}) are
\[ a_0=a_* \sin \eta, \quad \rho_0=\frac{3}{8\pi G}\,\frac{1}{a_*^2}\,\frac{1}{\sin^4 \eta}, \]
\begin{equation}
\theta_0=\frac{3\cos \eta}{a_*\sin^2\eta}, \label{56}
\end{equation}
where $a_*$ as before is given by Eq.~(\ref{33}). From
Eq.~(\ref{55}) we now have
\begin{equation}
\frac{d\rho_1}{d\eta}+\frac{4da_1}{d\eta}=24\pi Ga_*\sin \eta
\cos^2\eta, \label{57}
\end{equation}
which can be integrated from $\eta=\eta_{\rm in}$ onwards:
\begin{equation}\rho_1+4a_1=8\pi Ga_*(\cos^3\eta_{\rm
in}-\cos^3\eta).\label{58}
\end{equation}
 We have here assumed that $a_1=\rho_1=0$ at the initial instant
$t=t_{\rm in}$. With
\begin{equation}
\theta_1=\frac{da_1}{d\eta}\tan \eta \label{59}
\end{equation}
(cf. Eq.~(\ref{50})), we obtain from Eq.~(\ref{54}) a first order
differential equation for the scale factor correction:
\begin{equation}
\sin 2\eta\,\frac{da_1}{d\eta}+2(1+\cos^2\eta)a_1+8\pi Ga_*\cos^3
\eta=8\pi Ga_*\cos^3\eta_{\rm in}. \label{60}
\end{equation}
After some calculation we find the following solution, again
observing the initial conditions at $t=t_{\rm in}$:
\begin{equation}
a_1(\eta)=\frac{4\pi Ga_*}{\sin^2\eta}\left[ \cos^3\eta_{\rm
in}+\left( \frac{1}{4}\cos 2\eta-\cos^2\eta_{\rm
in}-\frac{1}{4}\cos 2\eta_{\rm in}\right)\cos \eta \right].
\label{61}
\end{equation}
Once the scale factor correction $a_1$ is known, the corresponding
density correction $\rho_1$ follows from Eq.~(\ref{58}).

From Eq.~(\ref{61}) it is apparent that the relative correction
$\zeta a_1$ to the scale factor is of order $4\pi G\zeta a_*$ or,
in dimensional units, $4\pi G\zeta a_*/c^3$. We here note that
$4\pi G\zeta/c^2=1/(3t_c)$, thus about $6.5 \times 10^{-30}~{\rm
s}^{-1}$ according to Eq.~(\ref{47}), which in dimensional form
reads $t_c^{-1}=12\pi G \zeta/c^2$. The quantity $a_*$, according
to Eq.~(\ref{56}), is the maximum value of the scale factor in a
nonviscous $k=1$ theory. Let us put $a_*$ equal to the commonly
accepted value of the radius of the universe, {\it i.e.,}
$a_*=10^{28}$ cm. Then we obtain $4\pi G\zeta a_*/c^3=2\times
10^{-12}$. The relative correction to the scale factor is thus in
this case very small. Physically, this is due to the fact that we
are considering an instant relatively late in the history of the
universe. If the bulk viscosity $\zeta$ were higher at earlier
times, or, if there were an {\it anisotropic} stage present in the
early universe at which the enormously higher {\it shear}
viscosity would come into play \cite{brevik94}, then the effect
 would be significantly enhanced.

 \section{Remarks on shear viscosity and the holographic entropy bound}

 We round off this article by making some remarks on the influence
 from shear viscosity in cosmology. The shear viscosity concept,
 as mentioned in the Introduction, has received considerable
 attention recently \cite{weinberg04,kovtun03,karch03,brevik04}.
 What we shall focus attention on here, is the suggestion of
 Kovtun {\it et al.} \cite{kovtun03} that there exists a
 cosmological universal lower bound on the ratio $\eta /s$, $\eta$
 denoting as usual the shear viscosity and $s$ the entropy per
 unit volume. The mentioned authors are concerned with the
 infrared properties of theories whose gravity duals contain a
 black brane with a nonvanishing Hawking temperature, the point
 being that the
infrared behavior is governed by hydrodynamical laws. If we for
definiteness consider a stack of $N$ non-extremal D3 branes in
type IIB supergravity, the metric near the horizon is given by
\begin{equation}
ds^2= \frac{r^2}{R^2}[-f(r)dt^2+dx^2+dy^2+dz^2]+
\frac{R^2}{r^2f(r)}dr^2+R^2d\Omega_5^2, \label{62}
\end{equation}
where $R\propto N^{1/4}$ is a constant, and $f(r)=1-r_0^4/r^4$
with $r_0$ being the horizon. The Hawking temperature of this
metric is $T=r_0/\pi R^2$, and $\eta$ and $s$ are given by
\begin{equation}
\eta=\frac{1}{8}\pi N^2T^3, \quad s=\frac{1}{2}\pi^2N^2T^3.
\label{63}
\end{equation}
Thus, when using hereafter  dimensional notation,
\begin{equation}
\frac{\eta}{s}=\frac{\hbar}{4\pi k_B}=6.08 \times 10^{-13}~{\rm
K\,s}. \label{64}
\end{equation}
The conjecture of Kovtun {\it et al.} is that the value in
Eq.~(\ref{64}) is a {\it lower bound} for $\eta/s$. Since this
bound does not involve the speed of light, the authors even
conjecture that this bound exists for all systems, including
non-relativistic ones. The idea has recently been further
elaborated in \cite{karch03}, arguing that the bound follows from
the generalized covariant entropy bound.

Let us check the proposed bound, by considering an example taken
from the plasma era in the early universe. We choose the same
instant as in subsection 4.1, namely $t_{in}=1000$ s,
corresponding to $n \simeq 10^{19}\; {\rm cm^{-3}}$ (subscript
"in" here omitted), $T\simeq 4\times 10^8$ K, whereas the energy
density is $\rho c^2=a_rT^4$, where $a_r=\pi^2
k_B^4/(15\hbar^3c^3)=7.56\times 10^{-15}\; {\rm
erg\,cm^{-3}\,K^{-4}}$ is the radiation constant. The viscosity
coefficients - whose existence is due to the imperfectness of
thermal equilibrium - can be calculated from relativistic kinetic
theory. Let $x=m_ec^2/k_BT$ be the ratio between electron rest
mass and thermal energy; when $x \gg 1$ it is convenient to use
the polynomial approximations
 \cite{caderni77} (cf. also \cite{brevik94}) for the evaluation of
$\eta$ and  $\zeta$:
\begin{equation}
\eta=\frac{5m_e^6 \,c^8\zeta(3)}{9\pi^3\hbar^3 \,e^4 \,n}x^{-4},
\quad \zeta=\frac{\pi c^2\hbar^3 n}{256\, e^4\zeta(3)}x^3,
\label{66}
\end{equation}
$\zeta(3) =1.202$ being the Riemann zeta function. At $T=4\times
10^8$ K one has $x=14.8$, leading to
\begin{equation}
\eta=2.8\times 10^{14}\; {\rm g\,cm^{-1}\,s^{-1}}, \label{67}
\end{equation}
which is enormously greater than $\zeta$ given by Eq.~(\ref{44}).
We note that both $\eta$ and $\zeta$  contain $\hbar$.

The entropy density, in view of the radiation dominance, is given
by
\begin{equation}
s=\frac{4}{3}a_rT^3=6.45\times 10^{11}\; {\rm
erg\,cm^{-3}\,K^{-1}}, \label{68}
\end{equation}
and so
\begin{equation}
\frac{\eta}{s}=435\; {\rm K\,s}. \label{69}
\end{equation}
This value is obviously much greater than the proposed lower bound
given by Eq.~(\ref{64}).

So far, we assumed a radiation dominated universe. If the state
equation is instead taken as $p=(\gamma -1)\rho c^2$, we obtain
analogously
\begin{equation}
\frac{\eta}{s}=\frac{578}{\gamma}\;\, {\rm K\,s}, \label{70}
\end{equation}
agreeing with Eq.~(\ref{69}) when $\gamma =4/3$.

Other cases analyzed in \cite{brevik04}, taken under widely
different physical conditions, turned out to give values of $\eta
/s$ of roughly the same order of magnitude as above.

\vspace{1cm} {\bf Acknowledgment}

\bigskip

I thank Professor Sergei Odintsov for valuable information.

\end{document}